\documentclass[twocolumn,aps,prl,showpacs,superscriptaddress]{revtex4}
\usepackage[dvips]{graphicx}
\usepackage{bm}
\begin{document}
\title{Lattice model calculations on aqueous acetone and tetramethyl urea}
\author{L. V.~Elnikova$^{a,b}$}
\affiliation{$^a$ Research Institute for Solid State Physics and Optics
Budapest 1525, POB 49, Hungary\\$^b$ A. I. Alikhanov Institute for Theoretical and Experimental Physics, \\
25, B. Cheremushkinskaya st., Moscow 117218, Russian Federation}

\date{\today}

\begin{abstract}

Self-organization of aqueous acetone and tetramethylurea, caused by lyotropic phase transformations, is studied in frames of the Ising model. Using lattice Monte Carlo simulations, the concentration dependent mixing schemes in these solutions are confirmed and the molar fractions corresponding to the transitions between the mesophases of these mixing schemes are theoretically calculated. 
\end{abstract}

%Keywords:

\maketitle

\section{Introduction}
In our study, we select two mixtures, aqueous solutions
of acetone and tetramethyl urea, where hydrogen bonding or the longer range dipole-dipole coupling plays an illustrative role in revealing structure properties of water.

Both these co-solvents are miscible with water over the entire
concentration range and at all temperatures between $RT$ ($R$ is the gas constant) and boiling
points. There is no lyotropic micellar phases, as the co-solvent
molecules are rather small and globular in shape. Still, according
to thermodynamic data, three composition regions can be identified
in these mixtures, in each of which the mixing scheme is qualitatively different from those in other regions. The structures and interactions within these regions are certainly defined by the features of the
constituent molecules, and can be roughly expressed in the following way.

In the region of low concentration of the co-solvent, a small number of AC molecules are bound together by C=O dipole attraction more tightly than in pure acetone, and for TMU solution, molecules are bound
together by $H=CH_3$ dipole attraction, respectively.
In solution, such a cluster enhances the hydrogen bond network of $H_2O$ in its immediate vicinity, with concomitant reduction in the hydrogen bond probability of bulk $H_2O$ away from solute. So, such a regions were introduced as concerning to Mixing Scheme I \cite{Koga1}. In the most surfactant-rich region, AC/TMU molecules form clusters in exactly the same manner as in pure liquid AC/TMU, in which AC/TMU molecules are bound together not only by $C=O$/$H=CH_3$ dipole but also van der Waals interaction. To the clusters of pure AC/TMU, $H_2O$ molecules interact as a single molecule. Hence, there is no hydrogen bond network of liquid $H_2O$. This is Mixing Scheme III \cite{Koga1}. Mixing Scheme II is operative, in which the solution consists of two kinds of clusters each reminiscent to I and III. The transition from Mixing Scheme I to II is associated with an anomaly in a thermodynamic quantity, and it occurs in a narrow mole fraction range.

For TMU, the effect of $H_2O$ was studied \cite{ref3}. TMU molecules directly participate in
forming the hydrogen bond network of $H_2O$, presumably via H donor on $NH_2$ and H acceptor on $C=O$,
up to almost the saturation. Hence, the hydrogen bond connectivity is not altered by the presence 
of urea.
However, the degree of fluctuations in entropy and volume, characteristic of liquid H$_2$O,
 is reduced \cite{ref3}.

There was found \cite{Katya}, that in water, two relaxation processes, differing only in activation entropy, may be due to two types of short-range order among molecules linked together by hydrogen bonds. Reorientation of the molecules occurs in structural defects, which allow the orientation of the molecules to change after the rupture a single hydrogen bond. The are two independent relaxation processes correspond to that possible types of defects. 

In this work, we compare the behaviors of certain thermodynamical values (an average energy 
and heat capacity) of aqueous AC and TMU in frames of statistical
analysis, which is useful for interpretation or confirmation of the
hypotheses on mesomorphism \cite{Koga1}. In our modeling, we involve
the Ising lattice version of the free energy of the systems, which was
carefully studied in respect to numerical Monte Carlo methods \cite{Allen}.
In particular, binary water-surfactant systems are quite enough described (\cite{Widom0}-\cite{4547}) via the low-temperature, the mean-field, and other approximations, and which are able to predict a majority mesomorphic states in aqueous solutions and can be tested numerically.

\section{Ising model and lattice Monte Carlo simulations}
The ability to formation of one or another space geometry of phases
depends including on a geometry and size of surfactant molecules, i.e. on its volume fraction in a solution. Chemical properties of heads and tails of surfactants, as well as their configuration thereupon, lead to formation of surfaces in solutions. The mesomorphism of aqueous AC and TMU also obeys universal phase behavior in lyotropic transitions, described in \cite{4547}, but a micellar degeneracy and other of these systems is outside of our report.

The presence of some experimental values of chemical potentials of acetone at molar fraction
 $0.1<x<0.16$, and for TMU at $0<x<0.03$ \cite{Koga1}, and in remainder ranges, calculated by 
the Boissonnas method, which allow as to perform a calculations in the lattice model. Through 
there are the estimating formulas \cite{Koga22, 8} on relation the between $\mu$ and $x$ 
in aqueous AC, or a while below,
 we omit it for the checking of validity of a lattice modeling.

Here, we apply the Ising gas lattice model \cite{4547}, where an each lattice site $u$ is 
assigned a spin variable $\sigma_u=0,1,-1$ of face-centered cubic lattice ($fcc$).
 Depending on the molar fraction of particles, distributed on a such lattice, we have a 
vacations, labeled as $-1$, water molecules ($0$), and surfactant's molecules (AC and TMU), 
which have assigned a variables $1$. We leave non-occupied lattice sites, because of arithmetic 
demand a total number of molecules in a system should be less or equal than a total number of 
lattice sites, on which we distribute our system; so we rarely have their coincidence. 
An orientational properties of surfactants are revealed via the coordinate variable
 $\mathbf{s_u}$ and $\mathbf{r}_{vu}$, where vectors $\mathbf{s_u}$ are assigned each s
ite of a lattice, and $\mathbf{r}_{vu}$ specify the directions from $fcc$ lattice site $u$ to $v$, 
for convenience, $|\mathbf{r}_{vu}|=1$. The Hamiltonian of the Ising model is:
\begin{widetext}
\begin{equation}\label{Hamiltonian}
H=-\frac{J_1}{4}\sum_{<uv>}\sigma_u(\sigma_u+1)\sigma_v(\sigma_v+1)-
\frac{J_2}{4}\sum_{<uv>}[(\sigma_u+1)(\mathbf{s}_u\cdot
\mathbf{r}_{uv})+(\sigma_v+1)(\mathbf{s}_v\cdot
\mathbf{r}_{vu})]-\frac{\mu_s}{2}\sum_u\sigma_u(1+\sigma_u),
\end{equation}
\end{widetext}
where $<uv>$ denotes a summation over all distinct pairs of
nearest neighbor sites $u$ and $v$. The term with coupling $J_1$ represents the isotropic interactions
between nearest-neighbor sites. The terms with coupling
constant $J_2$ describe the orientation-dependent
interaction between a AC/TMU molecule and a neighboring
water molecule. There is agreed that $J_2>0$ correspond to a surfactant
orientation with its head pointing toward a neighboring
water molecule. The last term gives the chemical
potential energy of the AC/TMU molecules. We remain here in the grand canonical ensemble, 
when the surfactant concentration is controlled by
the surfactant (AC/TMU) chemical potential $\mu_s$, rather than in
the canonical ensemble where the surfactant concentration is fixed.
Furthermore bellow, we will use a normalized values of excess
chemical potentials $\mu_s^{ex}$, instead of $\mu_s$. $\mu_s^{ex}$
were calculated and measured (as was reported in \cite{Koga1} and
references therein) depending on the surfactant mole fraction, they are only free of the linear 
terms. Such a substitution will not affect
on our average energy during Monte Carlo circles, since according to (\ref{Hamiltonian}), the 
reciprocal (first and second terms) and 'concentration' (third term) spin variables are summed 
up independently from each other.

To simplify the calculations, the Hamiltonian (\ref{Hamiltonian})
is normalized by the isotropic interactions $J_1$, and came to the
form:
\begin{widetext}
\begin{equation}\label{Hamiltonian2}
\widetilde{H}=-\frac{1}{4}\sum_{<uv>}\sigma_u(\sigma_u+1)\sigma_v(\sigma_v+1)-
\frac{\widetilde{j_2}}{4}\sum_{<uv>}[(\sigma_u+1)(\mathbf{s}_u\cdot
\mathbf{r}_{uv})+(\sigma_v+1)(\mathbf{s}_u\cdot
\mathbf{r}_{vu})]-\frac{\widetilde{\mu_s}}{2}\sum_u\sigma_u(1+\sigma_u)
\end{equation}
\end{widetext}
Basing on the results of low-temperature expansion \cite{4547}, we used the calculated relations 
between $\widetilde{j_2}$ and $\widetilde{\mu_s}$ at non-zero temperatures $t$ ($t=k_BT/J_1$, 
where $k_B$ is the Bolzmann factor, and $T$ is absolute temperature).
In this case, the excess chemical potentials be considered coinciding  with the standard ones.

Fortunately, the data of \cite{Koga1} and \cite{ref8}, cited in \cite{Koga1}, allow us to employ 
the known values of $\mu_s$, connected with its molar fraction at $T=25^oC$ as
\begin{equation}
\mu_i=RT\ln\frac{p_i}{x_ip^0_i},
\end{equation}
where $x_i$ is the mole fraction of $i$-component in solution and $p^0_i$ is the vapour pressure 
of pure $i$, $i$ means AC/TMU molecules, (see Table 1 and Table 2, squeezed from \cite{Koga1}).
\begin{table}
\caption{Some values of a mole fraction of aqueous acetone and chemical potentials at 25$^o$C 
\cite{Koga1}.}
\label{tab1}
\begin{tabular}{lcr}
$x_{AC}$  & $\mu_{AC}$ \\
1.000 & 0.000 & \\
0.770 & 0.223 & \\
0.580 & 0.580 & \\
0.350 & 1.200 & \\
0.223 & 2.300 & \\
0.175 & 3.000 & \\
0.100 & 4.000 & \\
0.000 & 5.000 & \\
\end{tabular}
\end{table}
\begin{table}
\caption{Some values of a mole fraction of aqueous TMU and chemical potentials at 25$^o$C \cite{Koga1}.}
\label{tab2}
\begin{tabular}{lcr}
$x_{TMU}$  & $\mu_{TMU}$ \\
0.900 & 0.000 & \\
0.750 & 0.100 & \\
0.600 & 0.300 & \\
0.500 & 0.400 & \\
0.320 & 0.500 & \\
0.080 & 0.550 & \\
0.070 & 0.650 & \\
0.020 & 0.950 & \\
0.015 & 1.200 & \\
\end{tabular}
\end{table}
So we performed our simulations, assigning the number of MC runs per a configuration 
10$^6$ and the number of thermalization steps equaling $10^5$. To minimize an energy of a 
configuration, we used the Metropolis algorithm building-in the Fortran program.

As a trial result, to confirm the presence of a transition states, corresponding to the Ising 
model \cite{4547} at high temperatures, we exhibit the temperature dependencies of specific 
heat (Figs. 1, 2) at $\widetilde{j_2}=1.5$, which was given from the expansion of the free 
energy, corresponding to the $fcc$ conditions of the task.

\begin{figure}\label{ct_ac}
\includegraphics*[width=70mm]{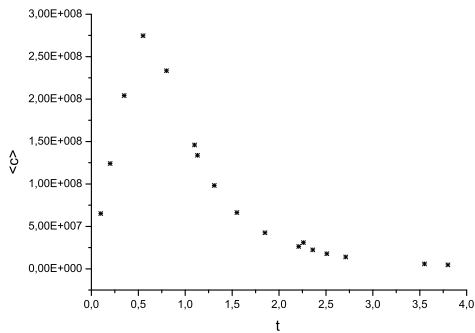}
\caption{\small The average heat capacity of aqueous acetone in dependence of temperature, 
obeying by MC simulations.}
\end{figure}

\begin{figure}\label{ct_tmu}
\includegraphics*[width=70mm]{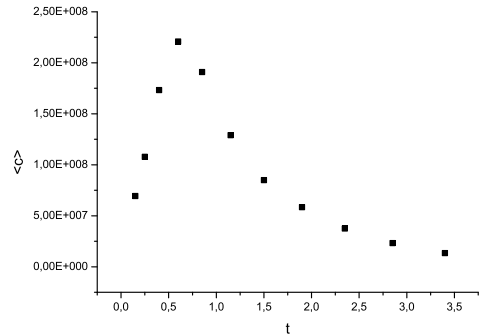}
\caption{\small The average heat capacity of aqueous TMU in dependence of temperature, obeying by 
MC simulations.}
\end{figure}

\section{Results and discussions}
In this section, we report and discuss our numerical results based on the model expressed by (1), 
in order to illustrate mesomorphism of aqueous AC/TMU.
\subsection{ Thermodynamics and structural schemes}  
From the Gibbs energy equation, the extensive thermodynamical quantities TS-pV remain after simulations, 
where $TS$ 
is the partial molar entropy, V is the volume of a given binary system. At the constant volume, $TS$
 behaves pro rata a free energy. From Figs. 3, 4, respective smoothness of the average energy, 
plotted for the aqueous 
AC and TMU systems, may be obvious.
In the real systems, the transition from Mixing Scheme I to II starts at point X at $x\simeq0.08$ 
(for AC and TMU) 
and ends at point Y, $x\simeq0.26$ \cite{Koga1}. However, according to the computer simulations, 
for both systems, the specific heat $<c>$ exhibits an absent of extremum between the points X and Y 
(Figs. 5, 6). In our MC tests, for aqueous AC, $<c>$ has a sharp maximum in the vicinity of the 
'II-III' 
transition (at  $x\simeq0.93$), and remains smooth in the region of the 'I-II' transition.
Fig. 7 shows the results of our MC tests, especially some typical disordered states at the Mixing 
Scheme 
II between the revealed transitions. Because of irrelevant number of empty lattice sites, 
only two molecular 
variables 0 and 1 are visible here. The picture is similar for aqueous TMU as well. 
Regions with clustered acetone 
and clustered water molecules are clearly seen.

The correlation functions (Fig. 8 and 9 for the systems of AC$-$water and TMU$-$water respectively) 
reflect an 
order parameter, analogous to the magnetization in a usual Ising magnetic. So, for AC$--$water, 
they show the 
temperature dependence taken at points of a molar fraction x corresponding to the mixing schemes. 

At low content of the solvent, 0.2, the order parameter reveals a non-smooth deep phase transition, 
which 
promptly reaches a plateau asymptotic with appreciable fluctuations. At an increase of concentration of 
AC, 
extremum indicated a phase transition is shifted to an increase of temperature, and an each curve 
becomes 
monotonic at a lower value of amplitude, than the foregoing one. 

For the AC−water system, our numerical results are comparable with such of \cite{13}, noted in 
\cite{Koga1}, 
for the concentration dependence of the excess partial molar volume in the 'I-II' transition point. 

For illustration, the results of the AC−AC volumetric interaction function are reported in \cite{Koga1}, 
we note, that
 the presence of water in our system as additional futures of the diagram of Fig. 8. 
Analogously for aqueous TMU (Fig. 9), the correlation functions demonstrate three different 
types of phase behavior, which correspond to three different mixing schemes.

\subsection{ Experimental evidence of clusterization for aqueous axcetone}
For the 'I-II' transition in AC, the MC results have qualitative correspondence with the statements 
of \cite{Katya}, obtained from the dielectric relaxation measurements and chemical analysis,
 that this is 
just the phase transition, connected with the second relaxation process at $x<$0.4, at a 
reorientation of 
the water molecule, involving a disruption of the inter-molecular hydrogen bonds in the dense 
structure. 
However, it is difficult to compare them directly because of different temperature, in which
 the chemical 
potentials were determined, and also thereupon that, in \cite{Koga1} 
and references therein, experimental excess 
chemical potentials were reported, while their total values were used in \cite{Katya}.
For the system 'water – acetone', 
coincidence of the calculated above and reported in \cite{Katya} data for minimum of 
the free energy at $x$ about 0.8 
was revealed.

\begin{figure}\label{ex_ac}
\includegraphics*[width=70mm]{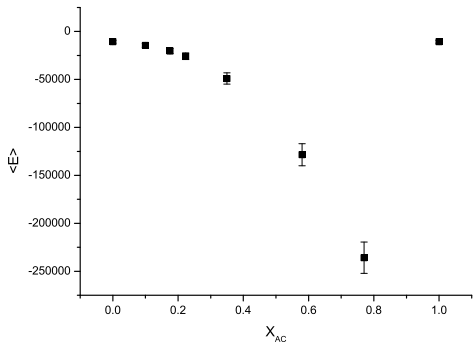}
\caption{\small MC results for the dependence of average energy on a molar fraction of aqueous acetone (AC), 
relative units.}
\end{figure}

\begin{figure}\label{ex_tmu}
\includegraphics*[width=70mm]{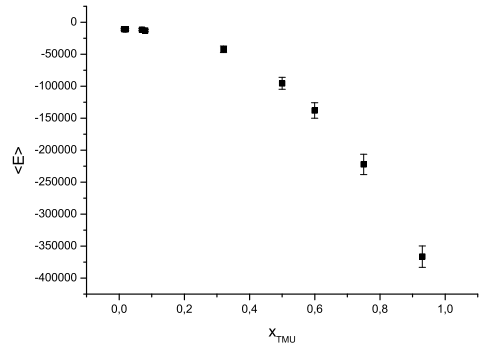}
\caption{\small MC results for the dependence of average energy (in relative units) on a molar fraction of 
aqueous TMU.}
\end{figure}

\begin{figure}\label{cx_ac}
\includegraphics*[width=70mm]{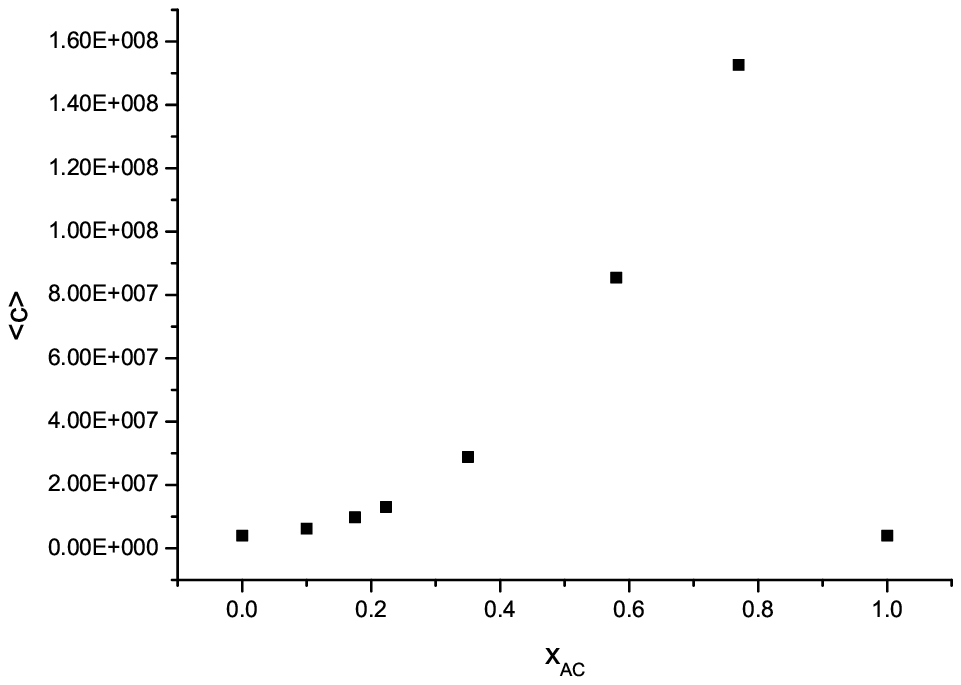}
\caption{\small MC results for the dependence of average heat capacity (in relative units) 
on a molar fraction of aqueous acetone (AC). Error bar is not shown.}
\end{figure}

\begin{figure}\label{cx_tmu}
\includegraphics*[width=70mm]{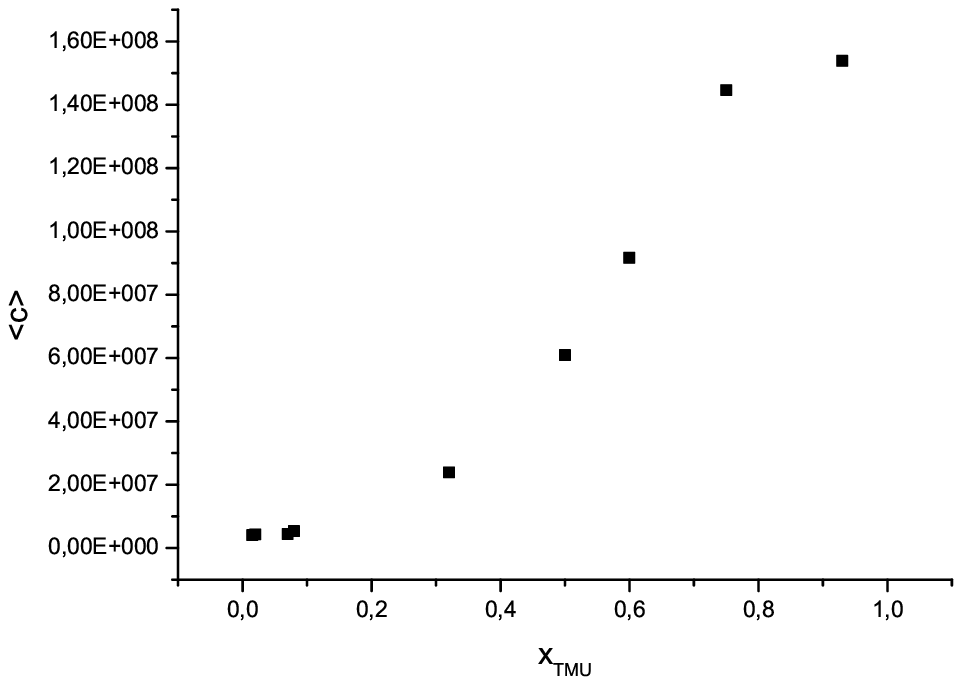}
\caption{\small MC results for the dependence of average heat capacity (in relative units) 
on a molar fraction of aqueous TMU. Error bar is not shown.}
\end{figure}

\begin{figure}\label{sresy}
\includegraphics*[width=70mm]{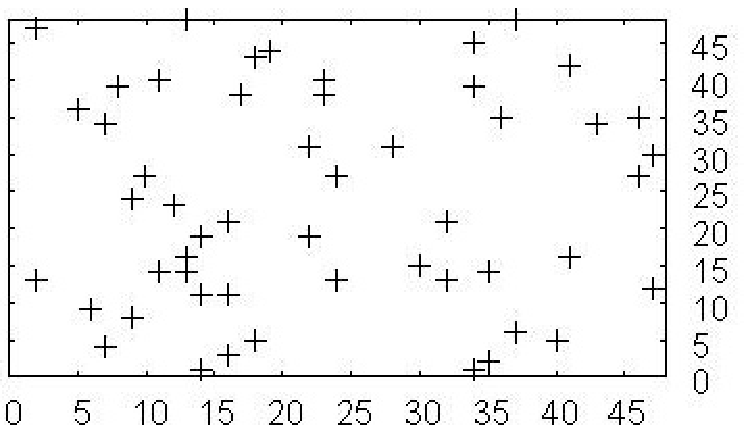}
\includegraphics*[width=70mm]{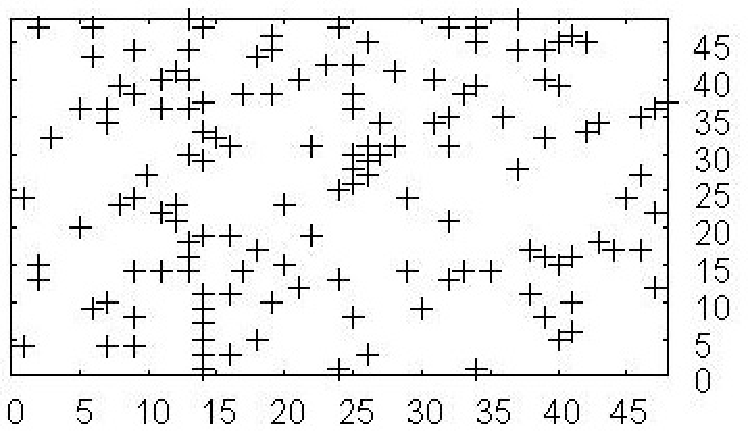}
\includegraphics*[width=70mm]{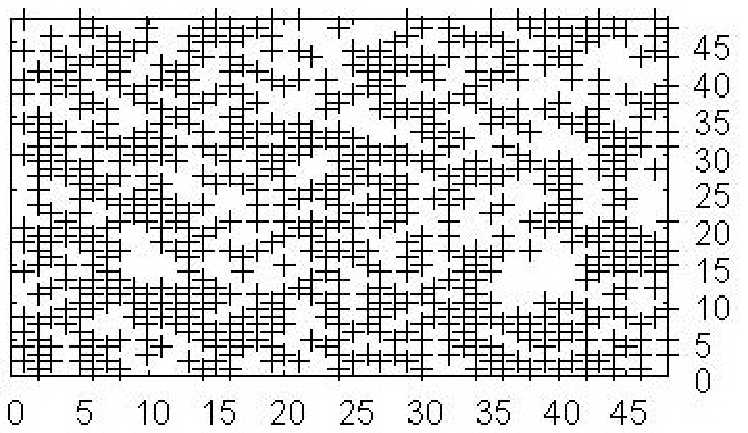}
\caption{\small A mid slices of MC lattice $48\times48\times48$ for AC molecule distribution in 
water, at molar fractions $x=0.100$, $x=0.223$, and $x=0.770$ respectively.}
\end{figure}
Improvement neutron small-angle scattering measurements and theoretical treatments via the 
angle-averaged radial distribution functions technique \cite{8} allow us to compare the hypothesis by 
Koga with the real states of molecular bonding in the case of the ACE$--$water system.

More exotic independent data on aqueous AC have been reached with positron annihilation spectroscopy 
\cite{15, 16}, where it was confirmed not only the mixing regions of the Koga’s studies, but the 
mechanism 
of clusterization was proposed in frames of atomic interactions (hydrogen bonding and so on).

\begin{figure}
\includegraphics*[width=70mm]{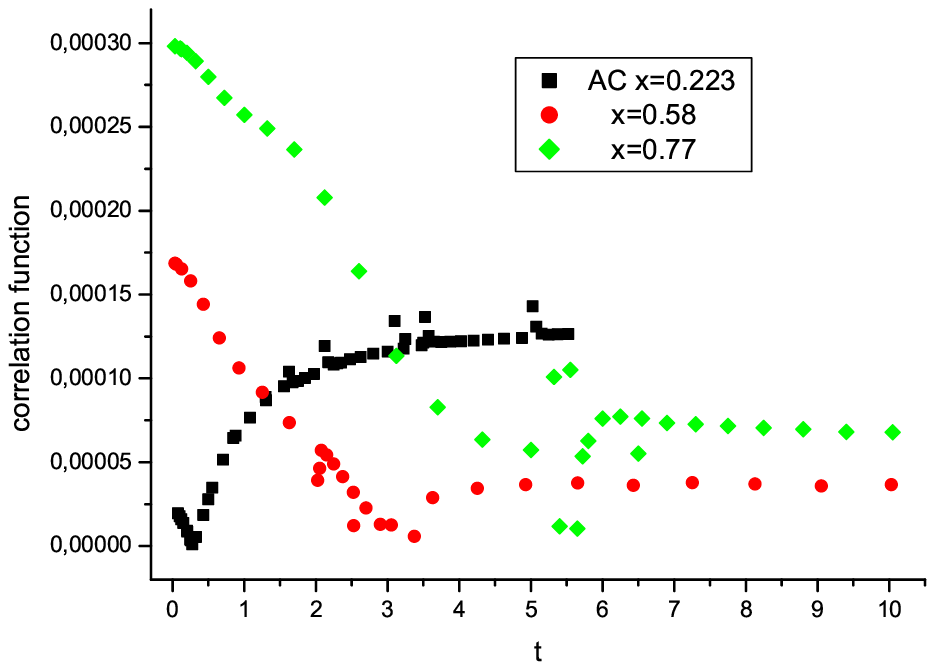}
\caption{\small MC results for the temperature dependence of the correlation functions of aqueous 
acetone in relative units. }
\end{figure}

\begin{figure}
\includegraphics*[width=70mm]{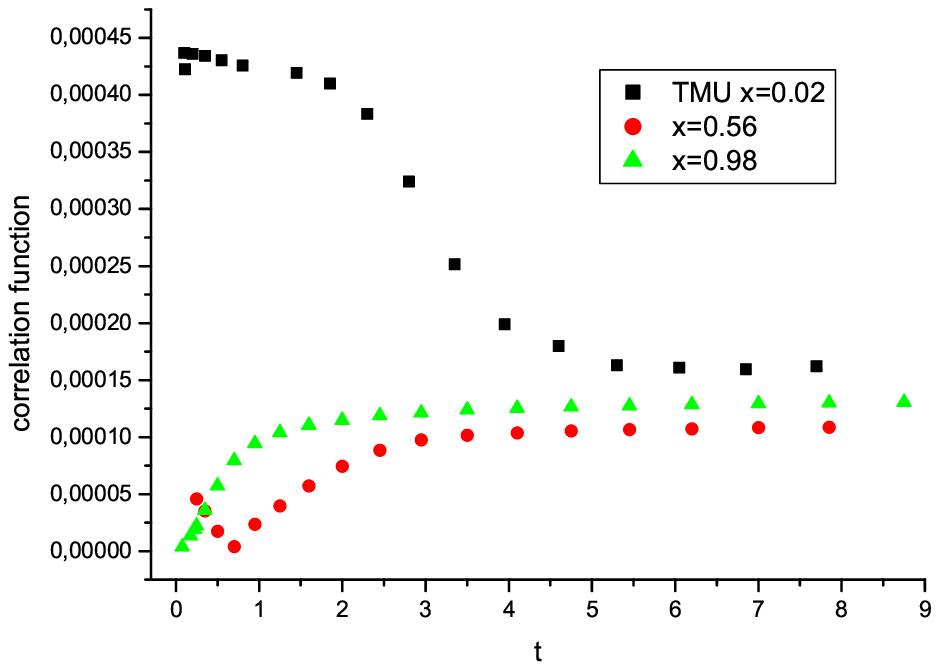}
\caption{\small  MC results for the temperature dependence of the correlation functions of aqueous
 TMU in relative units. }
\end{figure}

\section{Conclusion}
In polar aprotic solvents, the links with water are not strong, they are caused, in particular, by 
hydrogen bonding \cite{Katya}. This paper answers the question about the role of this bonding in 
mesomorphism 
of the whole system. With regards to above statistical analysis, one may confirm the data on 
dielectric 
relaxation \cite{Katya} 
by revealing of the phase transition at high AC (and TMU) concentrations, and at once, received an 
absent 
of coincidences at the low-concentration region. Viz., at the lyotropic structure rearrangement, there 
is 
not a distinct phase transition in a system completely. Therefore, the expected structures stay 
beyond the 
Ising model description in the region of high molar fraction of AC and TMU, even if it possesses more 
sensitivity to a transition. 
In this problem, it is important to separate a knowledge on pure water aggregates (hydrogen networks 
and cluster structures [3, 4]) and on binary water mixtures (aqueous AC or TMU), comparison the 
behavior 
of hydrogen bonding in water with that in solutions may stimulate further advance in the theory of water. 
Other aqueous aprotic solvents might be also used to contribute such data.

\textit{Acknowledgements.} The author thanks Balassi
Institute for the schoolarship financial supporting and the Neutron 
Spectroscopy Department of Research Institute for Solid State Physics and Optics of the Hungarian 
Academy of Sciences for
hospitality, and Dr. L. Alm\'{a}sy for useful discussion and his help in writing of the manuscript.

\end{document}